\documentclass[conference]{IEEEconf}

\input epsf
\usepackage{graphicx}
\usepackage{listings}
\usepackage{color}
\usepackage{multicol}
\usepackage{parcolumns}
\usepackage{float}
\usepackage{soul}
\usepackage{amsmath}
\usepackage{enumitem}
\definecolor{codegreen}{rgb}{0,0.6,0}
\definecolor{codegray}{rgb}{0.5,0.5,0.5}
\definecolor{codepurple}{rgb}{0.58,0,0.82}
\definecolor{backcolour}{rgb}{0.95,0.95,0.92}
\usepackage{lipsum}
\usepackage{url}
\usepackage[T1]{fontenc}
\usepackage{textcomp}
\usepackage{cite}

\renewcommand\thesection{\arabic{section}} % arabic numerals for the sections

\renewcommand\thesubsectiondis{\thesection.\arabic{subsection}}% arabic numerals for the subsections

\renewcommand\thesubsubsectiondis{\thesubsectiondis.\arabic{subsubsection}}% arabic numerals for the subsubsections

\begin{document}

\title{\small{Accepted by ISSSR 2024 (The 10th International Symposium on System Security, Safety, and Reliability) sponsored by the IEEE Reliability Society.}\\\textbf{\Large Specification and Enforcement of Activity Dependency  Policies using XACML\\}}

\author{Tanjila Mawla$^{1,*}$, Maanak Gupta$^{1}$, and Ravi Sandhu$^{2}$\\
	\normalsize $^{1}$Tennessee Tech University, Cookeville, Tennessee, USA\\
	\normalsize $^{2}$University of Texas at San Antonio, San Antonio, Texas, USA\\
	\normalsize tmawla42@tntech.edu, mgupta@tntech.edu, ravi.sandhu@utsa.edu\\
	\normalsize *corresponding author
}
%+++++++++++++++++++++++++++++++++++++++++++

%\specialpapernotice{(Invited Paper)}

\maketitle
\begin{abstract}
%In the era of technological advancements, the world is progressing towards a smart and futuristic community. 
The evolving smart and interconnected systems are designed to operate with minimal human intervention. Devices within these smart systems often engage in prolonged operations based on sensor data and contextual factors. Recently, an Activity-Centric Access Control (ACAC) model has been introduced to regulate these prolonged operations, referred to as activities, which undergo state changes over extended duration of time. Dependencies among different activities can influence and restrict the execution of one another, necessitating active and real-time monitoring of the dependencies between activities to prevent security violation. In the ACAC model, the activity dependencies, denoted as "D", is considered as a decision parameter for controlling a requested activity. These dependencies must be evaluated throughout all phases of an activity's life cycle.

To ensure the consistency of access control rules across diverse domains and applications, a standard policy language is essential. We propose a policy framework adapting the widely-used eXtensible Access Control Markup Language (XACML) , referred to as $\mathrm{XACML_{AD}}$,  to specify the activity dependency policies. %XACML is a reliable standard for enforcing access control policies in distributed yet connected domains. 
This work involves extending the syntax and semantics of XACML by introducing new elements to check dependent activities' states and handle state updates on dependent activities. In addition to the language extension, we present the enforcement architecture and data flow model of evaluating policies for activity dependencies. The integration of the proposed $\mathrm{XACML_{AD}}$ policy framework and the enforcement of the policies supports dependency evaluation, necessary updates and continuous enforcement of policies to control an activity throughout its life cycle. We implement the enforcement architecture exploiting the $\mathrm{XACML_{AD}}$ policy framework and discuss the performance evaluation results.
\end{abstract}
\IEEEoverridecommandlockouts
\vspace{1.5ex}

% for peerreview papers, inserts a page break and creates the second title.
% Will be ignored for other modes.
\IEEEpeerreviewmaketitle

\section{Introduction}
Security stands as a paramount element and a central focus for the creators and developers of systems and applications. Establishing a well-suited architecture for policy management is essential to ensure the integrity of interconnected systems. Going beyond conventional access control methods, security systems must be adaptable and foster interoperability across trusted domains \cite{lorch2003first}. Decisions regarding access to a system's resources by any requesting entity hinge upon criteria determined by business needs, requirements, and the designer's preferences. While traditional access control models concentrate on invariant attributes and their values to determine access requests' decision, modern smart systems with connected devices' operations still require context-based, fine-grained, flexible and comprehensive access control models. To address this, Gupta and Sandhu \cite{gupta2021towards} proposed the need for Activity-Centric Access Control (ACAC) model, followed by a mathematically grounded formal model by Mawla et al. to control long-lived activities that are performed by connected devices \cite{mawla2023acac_d, mawla2022bluesky}. The authors outlined access control decision components for such device activities including Authorizations (A), Obligations (B), Conditions (D) and Dependencies (D) on other activities. %The foundational concept of Activity-Centric Access Control (ACAC) was initially introduced by Gupta and Sandhu in 2021 \cite{gupta2021towards}. 
Since, activity performed by a device is a long-lived operation, to control such activities, the system needs continuous policy enforcement incorporating the decision parameters of an activity control.

XACML (eXtensible Access Control Markup Language) \cite{oasisopenEXtensibleAccess} stands out as a reliable standard for enforcing access control policies in distributed yet interconnected systems. Traditional access control models such as discretionary\cite{moffett1994specification} , mandatory\cite{fan2009mandatory}  and role-based (RBAC) \cite{ferraiolo2001proposed} access control policies have been specified using XACML syntax. The flexibility of attribute-based access control (ABAC) \cite{hu2013guide,gupta2022reachability} can be adapted by the policy enforcement using XACML. Continuous policy enforcement for grid computing \cite{feng2007resource}, usage control (UCON) \cite{katt2008general, colombo2010proposal, lazouski2012prototype} using XACML have been proposed in literature. These works reflect the extensions for XACML elements to accommodate access control policies for different security models.

The aim of this paper is to create a policy framework referred as $\mathrm{XACML_{AD}}$ for specifying the policy language to control smart system activities based on the activity dependencies \cite{mawla2023acac_d}. The dependencies (D) on activities is one of the decision parameters for activity control which needs to be evaluated at every phase of an activity's life cycle \cite{mawla2023acac_d}. Thus, continuous enforcement of activity dependencies policies is essential. The dependent activities must be in the desired states while a decision is made on the requested activity. If the current state of the dependent activities does not match with the desired state, the current state of a mutable activity will have to be updated before allowing the access. The update cannot be occurred if the dependent activity is immutable. In that case, the access will be denied. %The state update is only possible to occur on mutable dependent activities. 
Such update on dependent activities' states are not trivial as it may require check if there is any dependent of dependent activities leading to a chain of dependencies created where all dependent activities in the chain must have the desired states. The dependency evaluation along with updating current states require a formal language to specify the policies. We have analyzed the syntax and semantics of XACML 3.0 \cite{oasisopenEXtensibleAccess} policy specification and the related and nested entities of XACML 3.0 \cite{oasisopenXACMLV30} which we adapt for the policy specification of activity dependencies and continuous activity control. We also propose a new element for the state update actions by elaborating the necessary XACML syntax. We use the data flow model proposed in \cite{oasisopenEXtensibleAccess} making it compatible to the data flow of activity dependency policy evaluation during an activity access control. 

Rest of this paper is organized as follows. Section \ref{sec:background} provides the background of Activity-Centric Access Control (ACAC), activity dependencies and a review on XACML construct. Section \ref{sec:related-work} presents several state-of-the art works in continuous policy evaluation and XACML implementation. Section \ref{sec:running_policy} explains the policies which we cover in proposed $\mathrm{XACML_{AD}}$ policy framework. Further, the $\mathrm{XACML_{AD}}$ approach with XACML extension is described Section \ref{sec:xacmlad}. Section \ref{sec:enforcement} presents the architecture and data flow model of the policy evaluation using $\mathrm{XACML_{AD}}$ syntax. Section \ref{sec:implementation} shows the prototype implementation and the performance analysis of the proposed $\mathrm{XACML_{AD}}$ policy framework enforcement. Lastly, Section \ref{conclusion_future_work} summarizes the work and provides the future work direction.
\section{Background}
\label{sec:background}
\subsection{Activity-Centric Access Control} 
%Activity-Centric Access Control (ACAC) \cite{gupta2021towards,mawla2022bluesky,mawla2023acac_d} is a novel approach for controlling smart system devices' activities that are executed for a long duration of time based on the systems' needs. %Traditional access control models are designed to make decision of a requested right based on the subject's authorization to perform the right on the object considering different attribute values. 
%With the advancement of the technologies, the smart and connected systems are now offering fully or semi-automated environments where a large number of devices are connected and perform various operations with less human intervention. For instance, set of sensors collect data and provide input to the system for analysis, monitoring, and decision-making. On the other hand, there are different set of devices that include the sensors and perform long continuous activities based on the collected data from the sensors and other contextual conditions. These devices' long continuous operations are recognized as activities in the  ACAC model \cite{gupta2021towards,mawla2022bluesky,mawla2023acac_d}.

%Activities in smart systems are interconnected in smart systems such as smart farming, smart manufacturing systems. 
Today's smart systems (such as smart farming, smart manufacturing, etc.) are expected to work with less human intervention using the sensor data, evaluation of the environmental conditions and dependencies between multiple devices' operations. To meet these challenges, the activity-centric access control (ACAC) is introduced. Activity-Centric Access Control (ACAC) \cite{gupta2021towards,mawla2022bluesky,mawla2023acac_d} is a novel approach for controlling smart system devices' activities that are executed for a long duration of time based on the systems' needs.  ACAC combines the following aspects for access control decision: 1) considering dependencies among activities as a pivotal factor in activity control, and 2) ensuring continuous and active run-time enforcement of activity control parameters. In smart systems, such as smart farming, activities like plowing fields, pumping water, and spraying water are carried out by intelligent devices such as smart tractors, solar-powered smart pumps, and aerial drones, respectively. The proposed ACAC model encounters the following decision components to control activities at different decision time. 
\begin{itemize}
    \item \textbf{Authorization (A)}. Only authorized source requester can get access to the requested activity.
    \item \textbf{Obligations(B)}. Obligations are required one time actions that must be performed by the requesting source or any other subjects before an access being allowed to a requested activity.
    \item \textbf{Conditions (C)}. Environmental conditions must be evaluated before an access decision on an activity. For example, comparing the attribute values collected from the sensor data to pre-determined values should result in boolean "True" value in order to be satisfied as conditions.
    \item \textbf{Dependencies (D) on activities}. The relations (such as order of activities, concurrency) between the requested activity and other activities create dependencies that must be checked to ensure that the dependent activities have the desired current states while taking any decision (start, continue, hold) on the requested activity. We elaborate on this parameter in the next subsection. 
\end{itemize}

\subsection{Dependencies of Activities}

  In smart systems, the activities that have relationships with other activities in terms of execution order, concurrency, incompatibility, must ensure that these relations are maintained whenever access decisions on these activities are taken. For instance, when an activity is requested, two other activities need to be finished to ensure that the sequence of execution is maintained. These dependencies are checked in three different phases of the requested activity - pre, ongoing and post. Note that, each activity belongs to one of the states from inactive, dormant, aborted, running, revoked, hold or finished \cite{mawla2023acac_d}. 
  
  Once an activity is requested, the pre-dependent activities need to be in their desired states before allowing the requested activity to start. The continuity of the requested activity's execution depends on the fulfillment of the ongoing-dependencies. After a requested activity is finished or revoked (due to ongoing dependency violation), the post-dependent activities are evaluated. In these three cases, if the desired states of the dependent activities do not match with their current states, the activities must change their states to the desired ones to accommodate access decision on the requested activity. The dependent activities may depend on some other activities while changing their current state to the desired states and can form a chain of dependencies. In each level of the chain, the parent activity only can change the state if the child (dependent) activities are in their desired states. These recursive evaluation and update procedure as activity control parameter makes the ACAC model apart from other existing access control models. In this work, we specify activity dependency policies for evaluation, assuming the other decision parameters are checked and satisfied. %\hl{NEED TO MENTION ABOU THE STATES, AND REFER THE PAPER}

\subsection{XACML Review}
XACML (eXtensible Access Control Markup Language) \cite{oasisopenEXtensibleAccess} is a reliable and adaptable standard in terms of access control and security policy management. Access control in various applications and services, especially in distributed and web-based environments, can be flexible and extensible using XACML. XACML is developed and maintained by the OASIS (Organization for the Advancement of Structured Information Standards) consortium \cite{oasisopen}.

\begin{figure}[!htb]

\begin{lstlisting}[language=XML, caption={XACML Policy Construct}, linewidth=\columnwidth,frame=tlrb, label=lst-xacml_generic_structure,basicstyle=\footnotesize]
<PolicySet PolicySetId = ""
    PolicyCombiningAlgId = "">
    <Policy PolicyId = ""
        RuleCombiningAlgId = "">
        <Target>
            <Subjects>...</Subjects>
            <Resources>...</Resources>
            <Actions>...</Actions>
        </Target>
        <Rule RuleId = ""
            Effect = "">
            <Target>...</Target>
            <Condition>
                <Apply FunctionId="">...
                </Apply>
            </Condition>
        </Rule>
        <Obligations>
            <Obligation ObligationId = ""
                FulfillmentOn = "">
            </Obligation>
        </Obligations>
    </Policy>
</PolicySet>
\end{lstlisting}
\vspace{-4mm}
\end{figure}
According to the XACML policy specification language, policies are defined in XML format and specify rules for making access control decisions. A rule set or policy defines the structural organization of XACML. Rules may have a single condition or a set of conditions under which circumstances the decision from the rules will be applicable to the request.  In order to conclude with a final decision for the request, XACML supports combining algorithms that determine how multiple rules or policies are combined. The common combining algorithms are deny-overrides (if any decision is “Deny”, the result is “Deny”), permit-overrides (if any decision is “Permit”, the result is “Permit”), and first-applicable. For the `first-applicable', the policy evaluates the rules in the order that they are listed. In case of a specific rule, the outcome of the policy evaluation shall be `Permit', `Deny', or `Indeterminate', whichever is included as the effect of the rule if the target and the condition match. If a rule in the listed order does not match, the policy evaluation continues to the following rule. Obligations specify actions that must be performed after a final decision is provided for the policy evaluated, while advice provides suggestions that can influence the decision without being binding. Obligations are listed, including a `Fulfillment on' variable where the value can be `Permit' or `Deny' meaning that obligations must be fulfilled based on the policy decision of either permit or deny.

\textbf{Listing \ref{lst-xacml_generic_structure}} shows the XACML construct where a \texttt{<PolicySet>} is defined with a \texttt{`PolicySetId'} and a policy combining algorithm which is identified by the value of \texttt{`PolicyCombiningAlgId'}. The decision provided by a policy set is the result of combining the decisions from the child policies. A policy set contains 
several policies specified with the \texttt{<Policy>} tag, each of which defines the combining algorithm with a distinct \texttt{`PolicyId'} and a \texttt{`RuleCombiningAlgId'}. %\hl{WHERE IS DISCUSSION ABOUT POLICY COMBINE ALGO?} 
Each policy includes a \texttt{<Target>} element that specifies the subjects, resources, and actions to which the policy applies. Moreover, a  \texttt{<Policy>} contains particular rules, which are defined by a \texttt{`RuleId'} and an \texttt{`Effect'} representing the result of the rule if the target of the rule matches with the request and the condition inside the  \texttt{<Rule>} element is satisfied.

  XACML has an extensive set of functions. Functions are capable of operating on any set of attribute values and returning any type of value that the system supports. Additionally, functions can be nested, allowing for the creation of functions that takes the output of other functions as input in a complex hierarchy.  \texttt{<Apply>} element refers to the application of a function to its arguments, thus encoding the call of the function denoted with \texttt{`FunctionId'}. The  \texttt{<Obligations>} element offers further granularity, allowing the specification of single or multiple  \texttt{<Obligation>} elements, each specified with \texttt{`ObligationId'} and the \texttt{`FulfillmentOn'} attribute, to define the conditions triggering obligation fulfillment. 

\begin{figure*}[t]
    \centering
    \includegraphics[width=.7\textwidth,height=5cm]{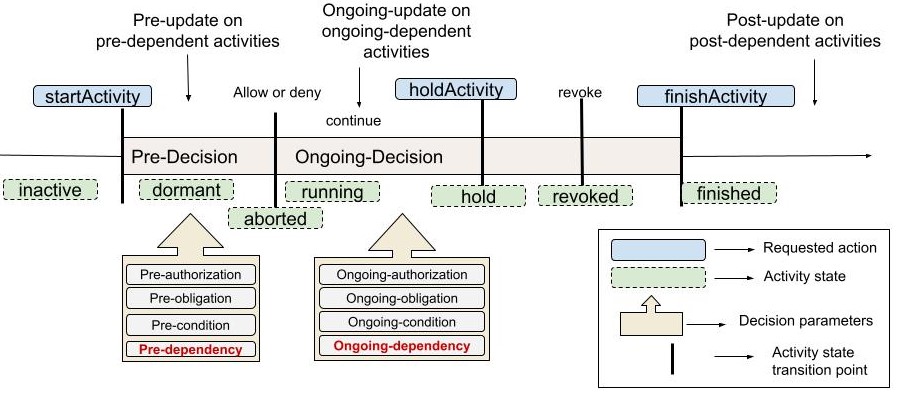}
    \centering
    %\vspace{-4mm}
    \caption{Activity-Centric Access Control (ACAC) with actions on a requested activity.}
    \label{fig-acac_with_policy_based_actions}
    \vspace{-4.5mm}
\end{figure*}
\section{Related Work}
\label{sec:related-work}
Several works \cite{riad2021adaptive, ashutosh2023xacml,dinh2012city,mourad2015sba, dallel2021secure} have been conducted on the the specification of the security policies across different domains utilizing the XACML \cite{oasisopenEXtensibleAccess} standard language. This section discusses literature related to continuous policy enforcement and the utilization of XACML in IoT (Internet of Things) and connected environments. 
Ashutosh et al. \cite{ashutosh2023xacml} introduce a data flow model based on Attribute-Based Access Control (ABAC), naming it eXtensible Access Control Markup Language for Mobility (XACML4M). Their model addresses research questions related to connected vehicle requirements, including Signal Access Control (SAC), Time-Based Access Control (TBAC), Location-Based Access Control (LBAC), and Frequency-Based Access Control (FBAC). The authors modify the standard XACML language by altering the data model, adding new data types to XACML policy, and introducing new components (e.g., Vehicle Data Environment (VDE) integrated with Policy Enforcement Point (PEP), time period data type, GeoLocation Provider, Polling Frequency Provider, Access Log Service) to the data flow model. However, practical implementation is hindered by lack of access to a real vehicle, limiting a comprehensive understanding of real-world effects. Dallel et al. \cite{dallel2021secure} propose a new XACML-based Access Control and Delegation (XACML-based ACD) mechanism, extending the XACML architecture by incorporating a Delegation Decision Point (DDP) to manage delegation control in smart building emergency situations. In order to solve security concerns originating from smart devices' interactions with the physical world and data processing, Fysarakis et al. introduce a Cross-domain Service Access Control for devices (XSACd) in \cite{fysarakis2018xsacd}. This framework combines Devices Profile for Web
Services (DPWS) for smart homes with XACML fine-grained access control. With an emphasis on the authorization elements of the smart devices, the platform-agnostic XSACd entities provide the smooth integration of access control across heterogeneous devices in smart residential settings with less user intervention.

%The update actions are performed inside the PDP as a part of the decision evaluation process. The actor of the update actions may or may not involve the human in the loop depending on the design of the system.

Several works have investigated continuous policy enforcement using Usage Control (UCON) \cite{zhang2008toward, hafner2008modeling, lazouski2012prototype, colombo2010proposal}. Hafner et al. \cite{hafner2008modeling} demonstrate the continuity of access decisions supported by policies and the decision engine in their SECTET-Framework for a healthcare system based on UCON \cite{park2004uconabc}. Colombo and co-authors \cite{colombo2010proposal} identify limitations in the current XACML standard in facilitating continuous usage control. They propose a U-XACML architecture, extending the syntax and semantics of XACML to incorporate continuity of access control and attribute mutability. Further, Lazouski et al. \cite{lazouski2012prototype} propose the U-XACML architecture with a mutable attribute retrieval model and a proof-of-concept implementation. While their work mentions the update of mutable attribute values in obligations, it lacks proper specification of the update procedure. In our work, we enforce activity continuity, clearly define extension of XACML syntax and semantics for the activity update procedure, and present experimental results showcasing improved time efficiency.
\section{Activity Execution Cycle and Dependency Policies}
\label{sec:running_policy}
%ACAC Dependency Evaluation and Policies

Figure \ref{fig-acac_with_policy_based_actions} illustrates the progress of an activity within it's life cycle. The green dotted boxes in the figure indicate the different states of an activity, namely `inactive', `dormant', `aborted', `running', `hold', `revoked', and `finished'. Each activity is associated with one of these states at any given time, as proposed in \cite{mawla2023acac_d}. The blue boxes indicate the requested actions by a requester (an user or system) and performed after evaluating the decision parameters. %The star symbol indicates a decision after the  parameters being evaluated. 
The black vertical line indicates the activity state transition point after the decision parameters are evaluated.

When an activity is in `inactive' state, a source initiates a $startActivity$ request, which transitions the activity to the `dormant' state. The figure includes the `Pre-Decision' phase, encompassing decisions to either allow or deny the requested activity to start. This determination relies on the source's authorization, conditions, obligations, and the fulfillment of dependencies on other activities, including necessary pre-updates on pre-dependent activities. If the pre-decision is `deny', the activity is aborted, else, the activity begins and transitions to `running' state. While the activity is in `running' state, the ongoing decisions to continue, revoke, hold, or finish the activity are evaluated. The system continuously checks the activity while in the `running' state. If the ongoing decision parameters are not satisfied, a `revoke' decision will be made, and the activity changes to `revoked' state.

During the execution of a requested activity A, another activity B with higher precedence may be requested by a source that could disrupt the ongoing requested activity. In such a scenario,  a $holdActivity$ request can be made by the system on activity A to allow activity B to complete its execution. The $finishActivity$ action is executed on the running requested activity without any activity dependency check. Following the completion of this action, the activity transitions to the `finished' state.

\iffalse
\hl{During the execution of a requested activity A, another activity B may be requested by a source that could disrupt the ongoing requested activity. In such a scenario, when a $holdActivity$ request is intercepted, the ongoing dependent activities of A are checked to see if B exists in that list.

%determine if a dependent activity is in a `dormant’ state and immutable. 

This situation indicates that the ongoing dependent activities that are immutable and in a dormant state, are emergency activities that need the running activity to be put on hold.

The system administrator must design policies and dependencies to ensure that emergency activities are listed as ongoing dependent activities. To continue the requested activity, these activities must be in states other than `dormant' or must be mutable. Otherwise, if the $holdActivity$ action is requested, it is performed on the running requested activity, causing the running activity to transition from `running' to hold'.}
The $finishActivity$ action is executed on the running requested activity \hl{without any activity dependency check. However, the requester's authorization, activity duration, and other environmental conditions must be verified, which is beyond the scope of this work. Following the completion of this action, the activity transitions to the `finished' state.}

\fi

\subsection{Running Policy Example}

Let us consider the following access control policies that are explained in natural language. Since an activity transitions from one state to another based on requested actions and evaluation of the decision parameters, it is essential to have policies based on which these actions are allowed or denied. In addition, the transition of a requested activity followed by an action performed and a decision taken on it, needs policy specification. In this paper, we work on the policy specification for activity dependency evaluation at different phase of an activity's life cycle assuming other parameters (authorization, obligation and condition) are checked and satisfied. %HOW DID YOU SELECT THESE SCENARIOS? WHY DO YOU THINK THEY ARE COMPREHENSIVE?
Enforcement of these policies accommodate the continuous policy evaluation for long-lived activities.

\begin{itemize}
    \item \textbf{Start activity without pre-dependent activities or with all pre-dependent activities in their desired states with or without state-updates:} An activity requested by an authorized user is allowed to be executed when all pre-dependent activities are in their desired states. For example, activity `plowing field' must be in the `finished' state before `sowing the seeds' is started. The current state of the 'plowing field' needs to be updated to `finished' if it is in a `running' state, when `sowing the seeds' is requested to be started.
    \item \textbf{Continue activity without ongoing-dependent activities or with all ongoing-dependent activities in their desired states with or without state-updates:} During the execution of an activity, it is imperative to evaluate its ongoing dependent activities to ascertain their adherence to the desired states. This evaluation should occur at defined intervals, preferably at small time intervals, say every 5 or 10 milliseconds. If the ongoing dependent activities are not in their desired states and are unable to update their states, the requested and running activity will be revoked from execution.
    % \item \textbf{Hold activity for accommodating certain temporary or emergency activities that are requested.} 
    % During the execution of a requested activity, the hold request on that activity can be allowed if there is an activity (emergency) which is requested and cannot update the state due to the emergency situation. That means that the emergency activity is in a `dormant' state and it cannot update its state without having the running requested activity on hold. For example, when `spraying fertilizer' is being executed, if `pumping water' from the crop field is requested during an emergency situation for excessive rainfall, `spraying fertilizer' will go to `hold' state (if hold activity is requested) and let `pumping water' be started.
    \item \textbf{Activity control post update:} After a requested activity is finished, the post-dependent activities are evaluated to check whether they are in their desired states or not. If any post-dependent activity is not in the desired state, post-update takes place on this particular dependent activity.
\end{itemize}

\section{$\mathrm{XACML_{AD}}$ Framework: 
Syntax and Semantics}
\label{sec:xacmlad}
\begin{figure}[!htb]
    \centering
   \includegraphics[width=0.85\columnwidth, height=3in]{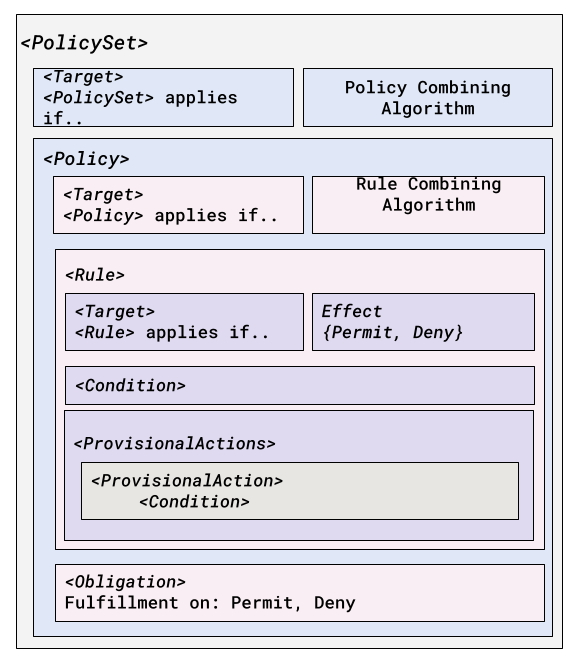}
    \centering    \caption{$\mathrm{XACML_{AD}}$ Constructs for Activity Dependencies Evaluation.}
    \label{fig-xacmlad_construct}
     \vspace{-4mm}
\end{figure}

%\hl{WHAT DO YOU MEAN BY \XAD APPROACH IN TITLE?}
Our goal is to express the policy language for the specification of activity-dependency policies to control a requested activity. We propose a $\mathrm{XACML_{AD}}$ policy framework which adapts existing XACML as well as proposes new XACML elements to accommodate the policy specification for activity dependency evaluation. We choose to utilize the XACML language since it is known for its widespread adoption in access control and ability to express application-independent language. The ability of XACML to support different domain requirements through arbitrary attributes encourages us to use it to define the dependencies on activities and write the policies for their evaluation. % \hl{NEED BRIEF MENTION OF WHAT IS \XAD?}
%\hl{I DON'T UNDERSTAND THIS...}
 Figure \ref{fig-xacmlad_construct} shows the constructs of our proposed $\mathrm{XACML_{AD}}$ policy framework used for the definition and evaluation of policies. 
%with rules under certain conditions. 
%The requested activity can have a set of pre-dependent activities (evaluated before starting the activity), ongoing-dependent activities (evaluated during the execution of the requested activity), and post-dependent activities (evaluated after the ongoing requested activity is revoked or finished). 
%In this work, we aim to evaluate the policy and rules to make a decision for the requested action on the requested activity based on the activity dependencies (D). %which is one of the decision parameters in our proposed Activity-Centric Access Control (ACAC) model. 

%We will first mention some assumptions before  discussion about the proposed extension of XACML in $\mathrm{XACML_{AD}}$ construct.%\\

%\textbf{Assumptions: }
We will first mention some assumptions before  discussion about the proposed extension of XACML in $\mathrm{XACML_{AD}}$ construct. In the ACAC model \cite{mawla2022bluesky,mawla2023acac_d}, the most suitable object is selected by the system when an activity is requested. The operation that triggers the object to start the requested activity is also retrieved from the system. The authorization of the source requester to access the requested activity and perform the corresponding operation on the object is checked according to the system security measures. In this paper, we assume the source requester is already authorized to access the requested activity and perform the corresponding operation. 
In addition, we assume the dependencies on other activities throughout all phases of the requested activity's life cycle (discussed in \cite{mawla2023acac_d}) are pre-defined without depending on a specific object.

%Dependency evaluation and the required updates on the dependent activities at different phases of activity make our proposed model novel and unique in terms of access control requirements.
The system must ensure that the dependent activities are in their desired states to allow an action on the requested activity. While specifying the dependency policies using XACML, we express policies for the possible actions that can be requested on an activity (note that, a requested activity is recognized as the resource in a $\mathrm{XACML_{AD}}$ request). The requested actions can be $startActivity$, $holdActivity$ and $finishActivity$. We express the policies mentioned in Section \ref{sec:running_policy} using the $\mathrm{XACML_{AD}}$ policy framework with the construct shown in Figure \ref{fig-xacmlad_construct}. We also have $continueActivity$ as an internal requested action on an activity to ensure that the policies are evaluated for ongoing-dependent activities to make a decision to continue or revoke the requested activity. If the matched action is `$startActivity$' for a policy specified within the policy set, all rules within that specific policy will be evaluated sequentially as listed until a decision found from any rule, since `first-applicable' rule combining algorithm is used. 

The rules within both the policies with matched action $startActivity$ and $continueActivity$ are similar but applicable to different sets of dependent activities (pre- and ongoing dependent activity sets, respectively). Note that, $continueActivity$ action is not directly requested by any requester. Rather this action is performed as an obligation after $startActivity$ action is permitted. We configure the rules with permit effect for the cases; (i) having no dependency; (ii) having dependent activities in their desired states; (iii) having dependent activities with update needed on mutable dependent activities with no dependent of dependent activities; and (iv) having dependent activities with update needed on mutable dependent activities with all dependent of dependent activities in their desired states. The only rule with deny effect is configured where the update on dependent activities is needed, but at least one of them is immutable and unable to change the state at the policy evaluation time. The current XACML language does not include the rules with any provisional action that is needed before allowing the requested action. Within the <Rule> element, we propose a <ProvisionalActions> element which includes single or several <provisionalAction> element including a <Condition> element as shown in Figure \ref{fig-xacmlad_construct}. The provisional action needed before allowing the requested actions to start or continue an activity is to update the states of the dependent activities. The condition is checked to ensure that these dependent activities are not in their desired states and are mutable at the evaluation time. Following the traditional XACML approach, we propose the XACML language extension for the provisional actions which shows similar XML-based language used in XACML. The <ProvisionalActions> element is a hierarchical element which aggregates a number of <ProvisionalAction> and <Condition> elements.

To represent the update on states of dependent activities as a requirement of allowing a requested action, we define a new element <ProvisionalActions> which consists of a collection of single <ProvisionalAction> elements:\\\\
<xs : element name = ``ProvisionalActions"\\
\hspace*{0.15in}type="$\mathrm{XACML_{AD}}$:ProvisionalActionsType"/>\\
<xs : complexType name = "ProvisionalActionsType">\\
	\hspace*{0.15in}<xs : sequence>\\
		\hspace*{0.3in}<xs : element ref = "$\mathrm{XACML_{AD}}$:ProvisionalAction"\\
			\hspace*{0.45in}maxOccurs = "unbounded"/>\\
	\hspace*{0.15in}</xs : sequence>\\
	\hspace*{0.15in}...\\
</xs : complexType>\\
\begin{figure*}[!htb]
    \centering
    \includegraphics[width=0.85\textwidth, height=2.5in]{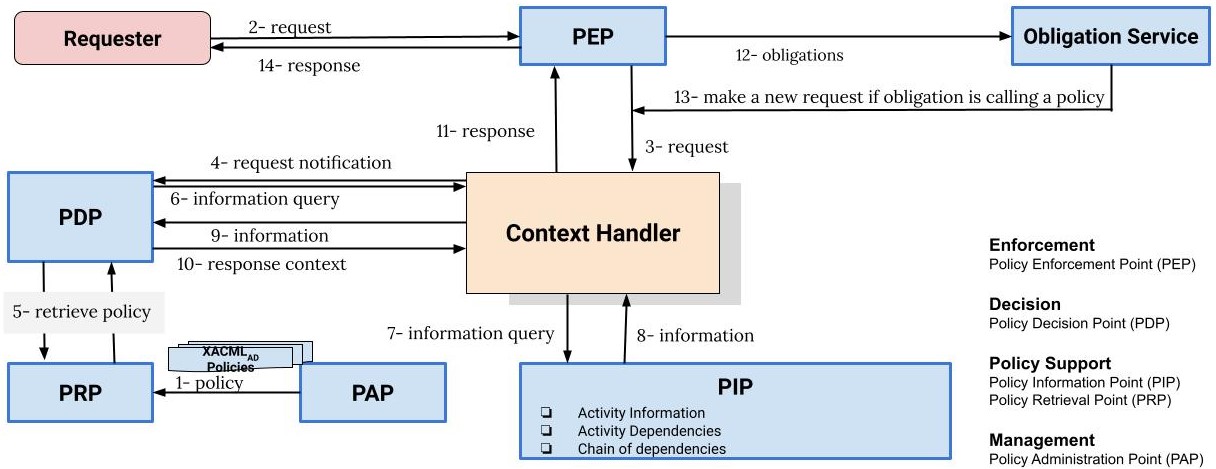}
    \centering
    %\vspace{-4mm}
    \caption{$\mathrm{XACML_{AD}}$ Enforcement Architecture and Data Flow.}
    \label{fig-enforcement-architecture}
    %\vspace{-8.5mm}
  
\end{figure*}
<ProvisionalAction> element defines each update on the states of dependent activities. The time of updates is defined by the value (pre, ongoing or post) of the attribute, `FulfillmentPhase'. The value of the attribute `ProvisionalAction' is always "Update" in our policies, since we use the element only for state update purpose.\\\\
<xs : element name = "ProvisionalAction"\\
\hspace*{0.15in}type="$\mathrm{XACML_{AD}}$:ProvisionalType"/>\\
<xs : complexType 
 name="ProvisionalActionType">\\
 \hspace*{0.15in}<xs : attribute name="FulfillmentPhase" type="xs : string"\\
\hspace*{0.3in}use="required"\\
\hspace*{0.15in}<xs : attribute name="ProvisionalAction" type="xs : string"\\
\hspace*{0.3in}use="required"/>\\
\hspace*{0.15in}<xs : element name = "Condition"\\
\hspace*{0.3in}maxOccurs = "unbounded"/>\\
	\hspace*{0.45in}...\\
</xs : complexType>\\

The existing <Condition> element in XACML is used within <ProvisionalAction> element to provide the flexibility of checking the conditions before performing a provisional action. In <ProvisionalActions> element, we also use the <ForAll> element to iterate over all items of a set of dependent activities and <ProvisionalAction> is exploited for each update performed on each dependent activity based on the condition written within the <Condition> element. The <ForAll> element is not shown in Figure \ref{fig-xacmlad_construct}. However, we use this element utilizing the structure of the XACML nested entity \cite{oasisopenXACMLV30}.

% \begin{figure*}[h]
%     \centering
%     \scalebox{.45}{
%     \includegraphics[]{Figures/XACMLAD_enforcement_architecture_v3.png}  }
%     \centering
%     %\vspace{-4mm}
%     \caption{XACMLAD Enforcement Architecture and Data Flow.}
%     \label{fig-enforcement-architecture}
%     %\vspace{-8.5mm}
  
% \end{figure*}

\section{$\mathrm{XACML_{AD}}$ Enforcement Architecture}
%Architecture and data flow supporting Enforcement of $\mathrm{XACML_{AD}}$ Policy Framework
\label{sec:enforcement}
In this section, we discuss the architecture and data flow which supports the enforcement of the $\mathrm{XACML_{AD}}$ policy framework. The proposed policy enforcement architecture starts with the interception of the request from the requester. The requested action on an activity can be $startActivity$, $continueActivity$ or $finishActivity$. $continueActivity$ is not requested by a requester, rather it is an action while evaluating the policies during an activity's runtime. After the request is intercepted, the decision is determined by evaluating the security policies in $\mathrm{XACML_{AD}}$. Figure \ref{fig-enforcement-architecture} shows the components of the policy enforcement architecture and the data flow from the request interception to providing the response, along with the fulfillment of obligations after response generation. The policy enforcement architecture consists of the following components:
\begin{itemize}[leftmargin=*]
    \item \textbf{Policy Administration Point (PAP)} creates and manages the policies and policy sets utilizing the $\mathrm{XACML_{AD}}$ policy framework (shown in Figure \ref{fig-xacmlad_construct}) and makes them available to the Policy Retrieval Point (PRP).
     \item \textbf{Policy Retrieval Point (PRP)} gets the policies and policy sets from the Policy Administration Point (PAP) and provides the necessary policies to the Policy Decision Point (PDP) which matches to the requested action in the $\mathrm{XACML_{AD}}$ request.
     \item \textbf{Policy Decision Point (PDP)} is responsible for the evaluation of the policies based on the request and the necessary information such as the current state of the requested and dependent activities, whether they are immutable or mutable and the dependencies including the chain of dependencies related to the requests.
     \item \textbf{Policy Enforcement Point (PEP)} intercepts the request from the requester and provides the response to the requester based on the decision (permit or deny) generated by the PDP.
     \item \textbf{Context Handler (CH)} constructs the request context from the native request provided by the PEP and notifies the PDP about the request with the necessary request information.
     \item \textbf{Policy Information Point (PIP)} stores the information that is needed for the policy evaluation by the PDP. PDP queries all the information needed for the policy evaluation to the context handler, and the context handler collects the information from PIP and sends it to the PDP for evaluation purposes. The PIP stores the information about the current state and mutability property of all activities, the dependencies of activities and chain of dependencies.
     \item \textbf{Obligation Service} fulfills the obligations provided by the PEP after making the activity permit or deny decisions. %The ACAC obligations are
      %Note that, the obligations mentioned in the ACAC model are different from these obligations. The obligations in ACAC is one of the prerequisites for activity decision. However, XACML obligations are performed after getting the permit or deny decision from the context handler. 
     In $\mathrm{XACML_{AD}}$ enforcement architecture, obligation service performs two types of obligations: i) updating the current state of the requested activity and ii) calling another policy such as the policy for $continueActivity$.
\end{itemize}

Adapting to the XACML \cite{oasisopenEXtensibleAccess} data flow model, Figure  \ref{fig-enforcement-architecture} illustrates the $\mathrm{XACML_{AD}}$ framework data flow sequence. 
%Figure \ref{fig-enforcement-architecture} follows the data flow sequence of the XACML \cite{oasisopenEXtensibleAccess} data flow model without loss of generality. 
The data flow towards enforcement of the architecture is described as follows:
\begin{enumerate}
    \item The PAP creates the $\mathrm{XACML_{AD}}$ policies. PAP makes the $\mathrm{XACML_{AD}}$ policies available to the Policy Retrieval Point (PRP). PRP configures and loads the policies and finds the appropriate policy when it is requested by PDP.
    \item The PEP intercepts the request from the requester.
    \item PEP sends the native request to the context handler.
    \item The context handler (CH) notifies the PDP about the current request along with sending the request information.
    \item The PDP retrieves the corresponding policy from the PRP from the policy set available in the PRP. Each policy is associated with a requested action in $\mathrm{XACML_{AD}}$ framework.
    \item PDP requests the necessary information associated with the retrieved policy to the CH. The necessary information includes the activity information such as the current state of an activity and whether it is immutable or mutable. Also, PDP asks for the information of the activity dependencies including the chain of dependencies needed for the policy evaluation.
    \item CH requests the PIP to provide the required information.
    \item The context handler collects it from the PIP to send back the information to the PDP.
    \item PDP collects the information from context handler for policy evaluation.
    \item Further, PDP evaluates the policy and prepares a decision for the requested action. At this phase, if there is any provisional action (update on dependent activities) needed to evaluate to a decision, PDP will perform it based on the information it has. The PDP sends the permit or deny decision after the policy evaluation as a response context to the context handler. \item The context handler prepares the final response, including the permit or deny decision, and sends the response to PEP.
    \item  PEP sends the obligations based on the decision in the response to the obligation service to let it fulfill the necessary obligations.
    \item If an obligation requires calling another policy, the obligation service makes a request containing the requested action and sends the request to the context handler. Further, steps 4-13 repeat.
    \item PEP sends the final response to the requester.
\end{enumerate}
\begin{figure*}

  \begin{minipage}{\textwidth}
     $<$PolicySet PolicySetId="1" PolicyCombiningAlgId= "only-one-applicable"$>$\\
    \hspace*{0.1in}$<$Policy PolicyId = "startActivityPolicy" RuleCombiningAlgId="first-applicable"$>$\\
        \hspace*{0.2in}$<$Target$>$$<$AccessSubject...$>$$<$Resource...$>$$<$Action$>$$<$ActionMatch$>$$</$Target$>$\\
        \hspace*{0.2in}$<$Rule RuleId="startActivityNoPreDep" Effect = "Permit"$>$$<$Condition...$>$$<$/Rule$>$\\
        \hspace*{0.2in}$<$Rule RuleId="startActivityWithPreDepNoUpdate" Effect = "Permit"$>$$<$Condition...$>$$<$/Rule$>$\\
        \hspace*{0.2in}$<$Rule RuleId="startActivityWithImmutablePreDepWithUpdateNeeded" Effect = "Deny"$>$$<$Condition...$>$$<$/Rule$>$\\
        \hspace*{0.2in}$<$Rule RuleId="startActivityWithPreDepUpdateNoDepOfDep" Effect = "Permit"$>$\\
            \hspace*{0.3in}$<$ProvisionalActions$>$...$<$/ProvisionalActions$>$$<$/Rule$>$\\
        \hspace*{0.2in}$<$Rule RuleId="startActivityWithPreDepUpdateWithDepOfDepNoUpdateNeeded" Effect = "Permit"$>$\\
        \hspace*{0.3in}$<$ProvisionalActions$>$...$<$/ProvisionalActions$>$$<$/Rule$>$\\
        \hspace*{0.2in}$<$ObligationExpressions$>$\\
        \hspace*{0.3in}$<$ObligationExpression ObligationId="updateRequestedActivityState" FulfillOn="Permit"$>$...$</$ObligationExpression$>$\\
            \hspace*{0.3in}$<$ObligationExpression ObligationId="call-continueActivityPolicy" FulfillOn="Permit"$>$...$</$ObligationExpression$>$\\
            \hspace*{0.3in}$<$ObligationExpression ObligationId="updateRequestedActivityState" FulfillOn="Deny"$>$...$</$ObligationExpression$>$\\
        \hspace*{0.2in}$<$/ObligationExpressions$>$$<$/Policy$>$\\
    \hspace*{0.1in}$<$Policy PolicyId = "continueActivityPolicy" RuleCombiningAlgId="first-applicable"$>$\\
    \hspace*{0.2in}$<$Target$>$$<$AccessSubject...$>$$<$Resource...$>$$<$Action$>$$<$ActionMatch$>$$</$Target$>$\\
    \hspace*{0.2in}$<$Rule RuleId="continueActivityNoOnDep" Effect = "Permit"$>$$<$Condition...$>$$<$/Rule$>$\\
    \hspace*{0.2in}$<$Rule RuleId="continueActivityWithOnDepNoUpdate" Effect = "Permit"$>$$<$Condition...$>$$<$/Rule$>$\\
    \hspace*{0.2in}$<$Rule RuleId="ongoingActivityWithImmutableOnDepWithUpdateNeeded" Effect = "Deny"$>$$<$Condition...$>$$<$/Rule$>$\\
    \hspace*{0.2in}$<$Rule RuleId="continueActivityWithOnDepUpdateNoDepOfDep" Effect = "Permit"$>$\\
        \hspace*{0.3in}$<$ProvisionalActions$>$...$<$/ProvisionalActions$>$$<$/Rule$>$\\
    \hspace*{0.2in}$<$Rule RuleId="continueActivityWithOnDepUpdateWithDepOfDepNoUpdateNeeded" Effect = "Permit"$>$\\
        \hspace*{0.3in}$<$ProvisionalActions$>$...$<$/ProvisionalActions$><$/Rule$>$\\
    \hspace*{0.2in}$<$ObligationExpressions$>$\\
        \hspace*{0.3in}$<$ObligationExpression ObligationId="call-continueActivityPolicy" FulfillOn="Permit"$>$...$<$/ObligationExpression$>$\\
        \hspace*{0.3in}$<$ObligationExpression ObligationId="updateRequestedActivityState" FulfillOn="Deny"$>$...$<$/ObligationExpression$>$\\
    \hspace*{0.2in}$<$/ObligationExpressions$><$/Policy$>$\\
\hspace*{0.1in}$<$Policy PolicyId = "finishActivityPolicy" RuleCombiningAlgId="permit-overrides"$>$\\
    \hspace*{0.2in}$<$Target$><$AccessSubject...$><$Resource...$><$Action$><$ActionMatch$>$$</$Target$>$\\
    \hspace*{0.2in}$<$Rule RuleId="finishActivityNoDependency" Effect = "Permit"$><$Condition$><$/Condition$><$/Rule$>$\\
    \hspace*{0.2in}$<$ObligationExpressions$>$\\
        \hspace*{0.3in}$<$ObligationExpressionObligationId="updateRequestedActivityState" FulfillOn="Permit"$>$...$<$/ObligationExpression$>$\\
        \hspace*{0.3in}$<$ObligationExpression ObligationId="call-postUpdatePolicy" FulfillOn="Permit"$>$...$<$/ObligationExpression$>$\\
    \hspace*{0.2in}$<$/ObligationExpressions$>$$<$/Policy$>$\\
\hspace*{0.1in}$<$Policy PolicyId = "postUpdatePolicy" RuleCombiningAlgId="first-applicable"$>$\\
    \hspace*{0.2in}$<$Target$>$$<$AccessSubject...$>$$<$Resource...$>$$<$Action$>$$<$ActionMatch$>$$</$Target$>$\\
    \hspace*{0.2in}$<$Rule RuleId="postUpdateNoPostDep" Effect = "Permit"$>$$<$Condition...$><$/Rule$>$\\
    \hspace*{0.2in}$<$Rule RuleId="postUpdateWithPostDepNoUpdate" Effect = "Permit"$><$Condition...$><$/Rule$>$\\
    \hspace*{0.2in}$<$Rule RuleId="postUpdateWithPostDepUpdateNoDepOfDep" Effect = "Permit"$>$\\
        \hspace*{0.3in}$<$ProvisionalActions...$>$$<$/Rule$>$\\
    \hspace*{0.2in}$<$Rule RuleId="postUpdateWithPostDepUpdateWithDepOfDepNoUpdateNeeded" Effect = "Permit"$>$\\
          \hspace*{0.3in}$<$ProvisionalActions...$>$$<$/Rule$>$\\
    \hspace*{0.2in}$<$ObligationExpressions$>$\\
        \hspace*{0.3in}$<$ObligationExpression ObligationId="updateRequestedActivityState"FulfillOn="Permit"$>$...$</$ObligationExpression$>$\\
    \hspace*{0.2in}$<$/ObligationExpressions$>$$<$/Policy$>$\\
% \hspace*{0.1in}$<$Policy PolicyId = "holdActivityPolicy"  RuleCombiningAlgId="permit-overrides"$>$\\
%     \hspace*{0.2in}$<$Target$>$$<$AccessSubject...$>$$<$Resource...$>$$<$Action$>$$<$ActionMatch$>$$</$Target$>$\\
%     \hspace*{0.2in}$<$Rule RuleId="holdActivityWithDormantImmutableOnDepAct" Effect = "Permit"$>$$<$Condition...$>$$<$/Rule$>$\\
%     \hspace*{0.2in}$<$ObligationExpressions$>$\\
%         \hspace*{0.3in}$<$ObligationExpression ObligationId="updateRunningRequestedActivityStateToFinished"FulfillOn="Permit"$>$...$<$/ObligationExpression$>$\\
%     \hspace*{0.2in}$<$/ObligationExpressions$>$$<$/Policy$>$\\
$<$/PolicySet$>$\\
\end{minipage}

\caption{Activity Dependency Evaluation and Update Policies using $\mathrm{XACML_{AD}}$Syntax}
\label{fig-overall_xacml_structure}
\end{figure*}

\begin{figure*}[htb]
    \centering
    \includegraphics[width=\textwidth]{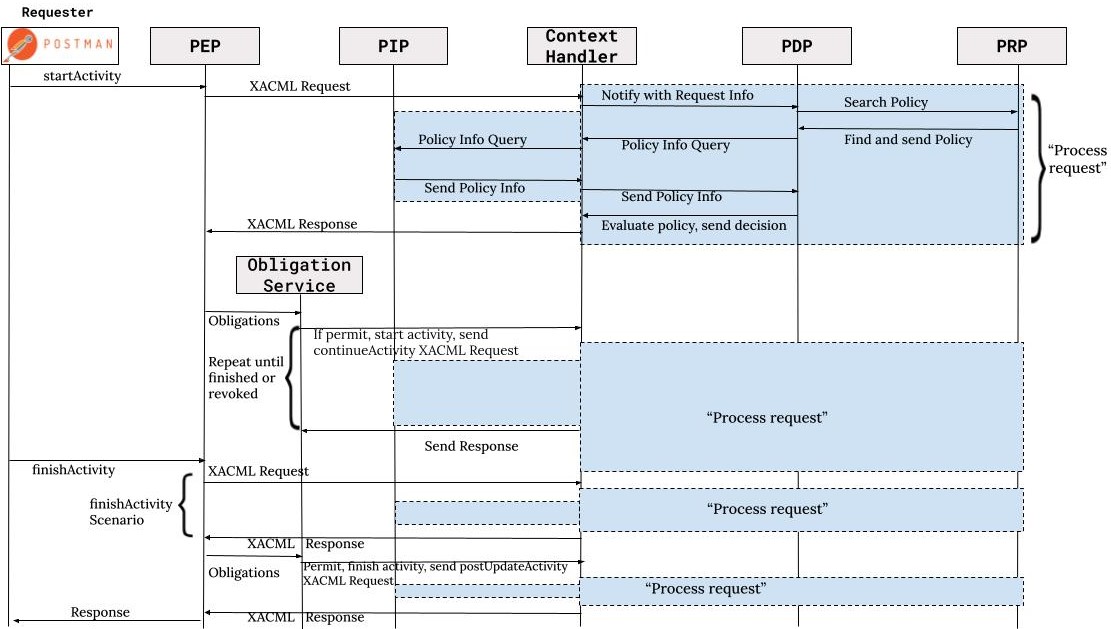}
    \centering
    %\vspace{-4mm}
    \caption{Sequence Diagram of the implementation of $\mathrm{XACML_{AD}}$ policy framework.}
    \label{fig-sequence_diagram}
    %\vspace{-8.5mm}
\end{figure*}

\section{Implementation of $\mathrm{XACML_{AD}}$ Policies}
\label{sec:implementation}
\sloppypar
We develop a policy engine for the implementation of the $\mathrm{XACML_{AD}}$ policy framework using Python programming language. The integration of the policy engine with the real systems depends on the designer's choice and the requirements of the system. For instance, the system administrator may decide to deploy the policy engine to a server and make the system interact with the server  by sending the access request through an API call. In this scenario, the standard API request-response communication protocols can be utilized. 

In our proposed framework, we use the original XACML constructs along with new elements as discussed in Section \ref{sec:xacmlad}. The overall policy structure in $\mathrm{XACML_{AD}}$ that utilizes XACML syntax for the activity dependency evaluation is shown in Figure \ref{fig-overall_xacml_structure}. We have listed the policies with their rules and obligations. For the sake of understanding, we only illustrate the relevant XACML elements to specify the $\mathrm{XACML_{AD}}$ policies.% Due to space constraints, the detailed and trivial $\mathrm{XACML_{AD}}$elements are not shown.

The obligations are used in two different purposes, both reflecting the tasks that need to be performed by the obligation service based on the final decision (permit or deny) by the PDP. The obligations include i) updating the current state of the requested activity based on the decision (reflected by `FulfillOn' attribute value), and ii) calling a policy. Each policy is defined using the `PolicyId' attribute. 
The policies called by obligation service include the policy defined with `PolicyId' value "continueActivityPolicy", to determine if the ongoing-dependencies are fulfilled to continue the activity execution. In our implementation, we can set-up the number of times and in what interval the evaluation of "continueActivityPolicy" will be performed until the activity is revoked or finished. Also, the policy with `PolicyId' value "postUpdatePolicy" is called after $finishActivity$ action is permitted.  

 We have a policy set which applies the policy combining algorithm "only-one-applicable" meaning that only one policy from the policy set needs to be evaluated to "Permit" or "Deny" decision. The appropriate policy is captured matching the target action by comparing the value of the attribute `action-id', to the value of `action-id' retrieved from the original $\mathrm{XACML_{AD}}$ request. The policies defined by the `PolicyId' values "startActivityPolicy", "continueActivityPolicy" and "postUpdatePolicy" include similar rules for dependency evaluation and update procedure of dependent activities' states as required. However, their evaluation time and obligations are different. We describe the rules included in the policy for `PolicyId' value "startActivityPolicy" which needs to be evaluated when the target `action-id' value is "startActivity". The policies capture the following requirements:

\begin{itemize}[leftmargin=*]
\item[--] All policies in the policy set follow the rule combining algorithm of "first-applicable" which denotes that the rule in the rule list whose condition is satisfied first is the ultimate rule providing the result of the policy evaluation.
\item[--] The rule specified with `RuleId'= "startActivityNoPreDep" is applicable if the requested activity does not have any pre-dependent activities resulting in "Permit" effect. The conditions are written and enclosed inside the <Condition> element.
\item[--] The rule specified with `RuleId'= "startActivityWithPreDepNoUpdate" is applicable when the requested activity has pre-dependent activities and all of them are in their desired states requiring no state update. The result of the evaluation is "Permit" in this rule.
\item[--] The rule specified with `RuleId'= "startActivityWithPreDepUpdateNoDepOfDep" is applicable when the requested activity has pre-dependent activities but all of them are not in their desired states. Thus, one or more pre-dependent activities need update from the current to the desired state. This rule is only applicable if their is no chain of dependencies for the pre-dependent activities which need to update their states. This rule's final effect is "Permit".
\item[--] The rule specified with `RuleId'=
"startActivityWithPreDepUpdateWithDepOfDepNoUpdateNeeded" is applicable if one or more pre-dependent activities need to update their states while having all of their dependent activities (dependent of dependent activities in a dependency chain) currently in the desired states. This rule's effect is "Permit".
\item[--] The rule specified with `RuleId'= "startActivityWithImmutableDepWithUpdateNeeded" is applicable if one or more pre-dependent activities or dependent of dependent activities in the dependency chain do not have same current and desired states while they are immutable at this moment of rule-checking. This immutability and need of update conflict with each other resulting the policy evaluation to a "Deny" Decision.
\end{itemize}

The policy with `PolicyId' value "continueActivityPolicy" evaluates ongoing dependent activities to provide the decision for the action $continueActivity$. The "finishActivityPolicy" always evaluates to "Permit" if the requester of the $finishActivity$ action is authorized to perform the action on the requested activity and the environmental conditions are met. This authorization process and environmental conditions check are out of scope of the paper. It is assumed that the requester is already authorized and all conditions are also satisfied. 
The policy with `PolicyId' value "postUpdatePolicy" evaluates the post-dependent activities after the $finishActivity$ action is performed. This policy is called as an obligation after $finishActivity$ action is permitted. This "postUpdatePolicy" has an obligation to update the requested activity's state from "finished" to "inactive". The obligations for updating the current state of the requested activity are identified by the value "updateRequestedActivityState" for  `ObligationId'. The values provided for the current state will be different at different phase. For clarification, after "startActivityPolicy" evaluates to "Permit", current state of the requested activity will be changed to "running". Similarly, if the "continueActivityPolicy" evaluates to "Deny", current state of the requested activity (which is running) will be updated to "revoked". This way, the framework is able to enforce the  continuous policy enforcement.

 %The obligations are included so that the requested activity's state can be updated after a policy is evaluated for a requested action following the decision provided by the PDP. 

%The XACMLAD policies are important for the continuous enforcement of activity control along with the classical authorization policies. The newly added element "<ProvisionalActions>" offers the flexibility of performing the required actions based on some conditions in specific rules under certain conditions. In this XACMLAD framework implementation, we exploit the enforcement architecture of the original XACML data flow model. We write the necessary obligations including calling for an existing policy while avoiding getting the request from external requester. This improves the security posture while giving the user least privilege. The insider threats who can be authorized users get a reduced chance to exploit the vulnerability of the system.

\subsection{Prototype Implementation}
The prototype implementation evaluates the activity dependency (D) decision parameter specified by $\mathrm{XACML_{AD}}$ policy framework. For the proof-of-concept purpose, 
%(D) without denying the significance of the evaluation of other decision parameters. Hence, 
we assume that other decision parameters are already evaluated and satisfied for the access decision.

We implement the functionalities of PEP, Context Handler (CH), PDP, PRP, PIP, PAP and the Obligation Service as shown in the enforcement architecture Figure \ref{fig-enforcement-architecture} using \texttt{Python} programming language. 
%We implemented the enforcement architecture (shown in Figure \ref{fig-enforcement-architecture}) for $\mathrm{XACML_{AD}}$policies using \texttt{Python} programming language. 
The JavaScript Object Notation (JSON) profile of XACML \cite{xacml-json-http-v1.1} currently handles the request and responses only. Considering the shortcoming of the JSON profile, we first write the $\mathrm{XACML_{AD}}$ policies, convert the policies to JSON format using Python xmlToDict\footnote{\url{https://www.askpython.com/python-modules/xmltodict-module}} library and write the dictionaries in JSON files.

PIP stores the necessary information about activity, dependencies of activities including chain of dependencies and provides to the context handler. Our policies are designed in a way that we do not write separate rules for the evaluation of each dependency. Rather we query for the dependencies, if PIP holds any, PDP is given those dependencies which are collected and sent by the CH. This gives the flexibility of changing the dependency requirements in any system.

%Depending on the domain and design of the request-response architecture, the requester can use different client interface and applications to request an action on an activity. 
In our implementation, we develop a python API (Application Programming Interface) in the PEP to intercept the activity action request that appears in form of HTTP request. We created a python API using Python Flask\footnote{\url{https://pypi.org/project/Flask/}} framework. The purpose is to get the request from any external service over HTTP protocol. However, we agree that the request-response architecture may vary depending on the application domain and the designer choice. We use Postman\footnote{\url{https://www.postman.com/}} to make the request to the API endpoint which resides in our python application. The requester uses the postman service to request an action (such as $startActivity$, $finishActivity$) on an activity. The python API in the PEP only accepts GET request and provides response in the JSON format using a response schema created in the implementation project. PEP communicates with the other modules to get the response for the request and sends the response for requested actions back to Postman. The continuity of an activity can also be evaluated reflecting the real environment where, for a specific duration of an activity, the continuity can be checked in certain intervals until the activity being revoked or finished.

Figure \ref{fig-sequence_diagram} shows a sequence diagram for the implementation of the $\mathrm{XACML_{AD}}$ policy framework for activity control. A $startActivity$ action is requested by the requester using postman and intercepted by the API residing in PEP. PEP sends the XACML request to the CH. CH notifies PDP about the request information, PDP retrieves the matched policy for the requested action from the request information and queries the information about the requested activity and its dependencies to the context handler (CH). CH collects the necessary information from PIP and sends the information to PDP. Further, PDP evaluates the associated policy and sends the decision to Context Handler. CH sends the XACML response to PEP for $startActivity$ action. PEP calls the obligation service to execute the obligations associated with the decision. If the decision is "permit", the obligation service changes the current state of the requested activity to "running". Further, the obligation service sends a request for $continueActivity$ action to the context handler. The "Process request" term indicated by the blue-colored portion repeats. "continueActivityPolicy" evaluation for $continueActivity$ action is repeated until the running and requested activity is revoked or finished. Later, a $finishActivity$ action request is sent to the PEP and the response is provided after the obligations are fulfilled. The obligations after finishing the activity includes changing the current state of the requested activity to "finished" and calling "postUpdatePolicy". For the evaluation of the "postUpdatePolicy", again "Process request" repeats and a final XACML Response is provided to the requester.

\begin{figure}[!t]
    \centering
\includegraphics[width=\columnwidth,height=2.55in]{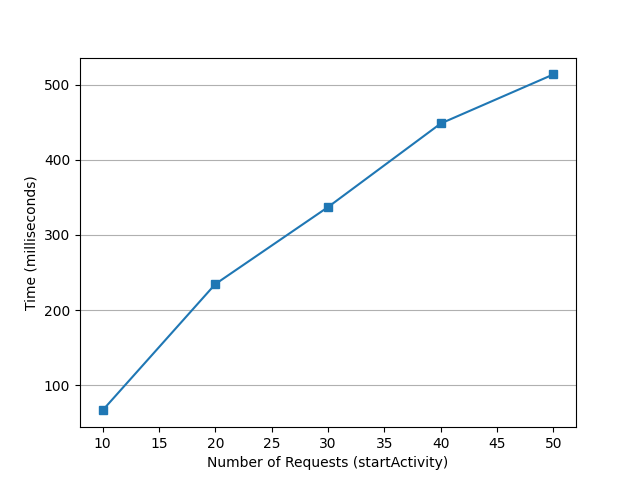}
    \caption{Execution time against the number of requests for $startActivity$ action.}
    \label{fig-performance_start}
     \vspace{-4mm}
\end{figure}
\subsection{Performance Evaluation}
We evaluated the performance of our $\mathrm{XACML_{AD}}$ policy framework implementation of using the server hosted in a local machine. The machine configuration includes the operating system of `Windows 11', programming language support of `Python 3.9' and memory of `Intel Core i7' with 1.7GHz processor and 16GB RAM.

\begin{figure}[!t]
    \centering
    \scalebox{.55}{
    \includegraphics[]{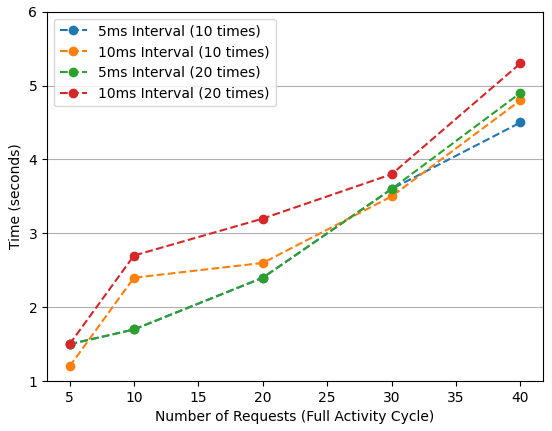}}
    \caption{Execution time against number of requests (full activity cycle).}
    \label{fig-performance_fullcycle1}
     \vspace{-4mm}
\end{figure}
\iffalse
\begin{figure}[!t]
    \centering
    \includegraphics[width=\columnwidth]{Figures/20times_5and10ms.png}
    \centering
    %\vspace{-4mm}
    \caption{Execution time against number of requests \hl{(full activity cycle with continuity evaluation 20 times )}.}
    \label{fig-performance_fullcycle2}
    %\vspace{-8.5mm}
\end{figure}
\fi
%We measured the performance of our proposed $\mathrm{XACML_{AD}}$implementation in the following criteria. 
We calculated the time consumption for the execution of a number of $startActivity$ requests on different activities. We initiate the request using postman where we ran a collection of 10, 20, 30, 40 and 50 requests for $startActivity$ action. PEP intercepts the request through the Python API and completes the evaluation process following the data flow model. In Figure \ref{fig-performance_start}, we show the execution time to perform $startActivity$ action on different number of requested activities. The execution time increases with the number of requests. We obtained that approximately 513 ms time is required for starting 50 new activities. The average time consumption per activity start is 10.26 ms. This result of execution time per $startActivity$ access to an activity is comparable to the average time (21.5 ms and 45.3 ms) per access evaluation in literature \cite{zhang2008toward,lazouski2012prototype}. Our enforcement model performs significantly better in terms of time-efficiency.  
Since activity is a long continuous operation performed for a duration of time, we also measured the execution time of the full activity cycle. The continuous dependency evaluation process takes variable amount of time based on the time interval of re-evaluation of ongoing-dependent activities and the duration of continuity of the activity. We also measured the response time from starting the $startActivity$ action request to the $finishActivity$ action request on different number of activities. Note that, postman can run a collection of requests sequentially which we used to measure the time for getting response including $startActivity$ and $finishActivity$ action request. On the other hand, the evaluation of ongoing-dependent activities and the required update processing is set up to be executed 10 times and 20 times with 5ms and 10ms intervals.

% \begin{figure}[!htb]
%     \centering
%     \includegraphics[width=\columnwidth]{Figures/holdActivityRequestInMilliseconds.png}
%     \centering
%     %\vspace{-4mm}
%     \caption{Execution time against number of requests for "$holdActivity$".}
%     \label{fig-performance_hold}
%     %\vspace{-8.5mm}

% \end{figure}
In Figure \ref{fig-performance_fullcycle1}, we show the execution time for full activity life cycle (start to finish) with $continueActivity$ action performed 10 and 20 times with 5 ms and 10 ms intervals. 
%The `blue' and `red' colors, respectively, indicate 5 ms and 10 ms intervals in the continuity re-evaluation. 
For continuity re-evaluation 10 times with 5 ms intervals, the average time for a full activity cycle is 112.5 ms while with 10 ms intervals this average execution time is 120 ms. We also observe that with $continueActivity$ action performed 20 times, the average execution time for full activity life cycle (start to finish) are 122.5 ms (with 5ms intervals) and 132.5 ms (with 10ms intervals). We conjecture from this observation that the number of time continuity evaluation performed can be increased according to the system needs while getting only milliseconds of time-difference.

We understand that real world smart systems involve hundreds and thousands of devices and activities that are interconnected. An extensive performance analysis with higher loads of activities and their dependencies is necessary while we believe that our proof-of-concept implementation is able to showcase the practical viability by implementing the  policy enforcement architecture for the extended XACML policy expressions.

% In Figure \ref{fig-performance_hold}, we also measured execution time against number of requests for `$holdActivity$' action to be performed on the running requested activities. The average time for holding an activity is 0.06ms. This fraction of milliseconds is needed just to confirm one condition of having at least a dormant and immutable ongoing dependent activity which indicates that an emergency activity is requested and waiting for execution by holding the running requested activity.

%\input{Sections/Related_work}
\section{Conclusion and future work}
\label{conclusion_future_work}
\sloppypar
This paper introduces a $\mathrm{XACML_{AD}}$ policy framework for specifying and implementing activity dependency policies for activity centric access control. %, in regards of making decisions to requested actions on smart system devices' activities. 
We use a widely adopted XACML policy language and extend its elements to express policies for the evaluation of dependent activity and updating the states based on the conditions. We also provide a $\mathrm{XACML_{AD}}$ policy enforcement architecture and the data flow model to enforce the policies for activity dependency evaluation for activity access decision. The enforcement architecture is constructed based on the data flow model outlined in the standard XACML 3.0. We use the data flow while incorporating an indirect communication between the Policy Decision Point (PDP) and Policy Administration Point (PAP) using a Policy Retrieval Point (PRP) component to identify the appropriately matched policy. In implementation of the enforcement architecture, we employ policies written in $\mathrm{XACML_{AD}}$. Our policies demonstrate expressive capabilities which along with the simplicity of the enforcement architecture contributes significantly to continuous policy evaluation in activity control. We also measure the performance of the implementation by sending requests (for actions on activities) over HTTP protocol. The experimental results exhibit promising time-efficient performance with varying numbers of requests.

%In this context, activities refer to prolonged operations performed by devices in smart and connected systems. 

%\hl{  
In our research, we conduct policy evaluation for two levels of activity dependencies within the dependency chain. It would be critical to enhance the $\mathrm{XACML_{AD}}$ policy framework to handle any number of dependent-activity levels. Our work covers the policy evaluation that includes state update actions based on conditions. 
The current XACML profile lacks constructs for recursive policy evaluation and does not provide syntax for state update actions. 
We leave the refinement of $\mathrm{XACML_{AD}}$ semantics for recursive updates in the dependency chain as a future direction for improvement. We believe that enhancing the $\mathrm{XACML_{AD}}$ framework in this way will effectively address the dependency of activities across large number of levels in the dependency chain.%}

\section*{Acknowledgement}
\vspace{-2mm}
This work is partially supported by NSF grants 2230609 and 1736209.

\bibliographystyle{unsrt}
\bibliography{bibliography}

\end{document}